\documentclass[12pt]{article}

\usepackage{amsmath,rotating,cite,graphics} 

\def\theequation{\arabic{section}.\arabic{equation}}

\newcommand{\mycaption}[2][kurz]{{\begin{center} 
\parbox{15cm}{{\bf \caption[#1]{\rm {#2}}}} \end{center} }}

\newcommand{\pb}[3]{{\parbox{#1 cm}{ \vspace{#2 cm} \begin{center} #3
\end{center}}}   
\vspace{#2 cm}}

\newcommand{\la}{{\mathcal L}}
\newcommand{\da}{^{\dagger}}

\newcommand{\lr}[1]{{ \left( \, #1 \, \right) }}
\newcommand{\lreckig}[1]{{ \left[ \, #1 \, \right] }}
\newcommand{\mr}[1]{{\mathrm{#1}}}

\newcommand{\Psibar}{\overline{\Psi}}

\begin{document}

\begin{flushright}
WUE-ITP-2003-025\\[-0.15cm] 
MC-TH-2003-07\\[-0.15cm]
hep-ph/0312186\\[-0.15cm]
December 2003
\end{flushright}

\begin{center}
{\Large {\bf Probing Minimal 5D Extensions of the}}\\[0.3cm]
{\Large {\bf Standard Model: From LEP to an $e^+e^-$ Linear Collider}}\\[1.4cm] 
{\large  Alexander M\"uck$^{\, a}$, Apostolos Pilaftsis$^{\, b}$ 
and Reinhold R\"uckl$^{\,a}$}\\[0.4cm]
$^a${\em Institut f\"ur Theoretische Physik und Astrophysik,
         Universit\"at W\"urzburg,\\ Am Hubland, 97074 W\"urzburg, 
         Germany}\\[0.2cm]
$^b${\em Department of Physics and Astronomy, University of Manchester,\\
         Manchester M13 9PL, United Kingdom}
\end{center}
\vskip1.cm   \centerline{\bf   ABSTRACT}   
\noindent
We derive new improved   constraints on the compactification  scale of
minimal 5-dimensional (5D) extensions of  the Standard Model (SM) from
electroweak and LEP2 data and estimate the reach of an $e^+e^-$ linear
collider such  as  TESLA.    Our  analysis  is  performed within   the
framework of  non-universal 5D  models, where some   of the gauge  and
Higgs fields propagate in the extra dimension,  while all fermions are
localized  on  a  $S^1/Z_2$    orbifold  fixed  point.   Carrying  out
simultaneous multi-parameter fits  of  the compactification  scale and
the SM parameters to the data, we obtain lower bounds on this scale in
the range between 4  and 6~TeV. These  fits also yield the correlation
of the compactification scale with the SM Higgs  mass.   Investigating
the  prospects  at  TESLA, we show that the so-called GigaZ option has 
the potential to improve these bounds by  about  a  factor 2 in almost
all 5D models.  Furthermore, at the center of  mass  energy of 800~GeV
and with an integrated luminosity of $10^3$~fb$^{-1}$, linear collider
experiments can  probe   compactification  scales up  to   20--30~TeV,
depending on the control of systematic errors.

\newpage

\setcounter{equation}{0}
\section{Introduction}
\indent

The  Kaluza and   Klein~(KK)   paradigm~\cite{KK} that our  world  may
realize more than four dimensions has been a central theme of the last
ten years~\cite{IA,EW,ADD,DDG}.   The additional dimensions have to be
sufficiently compact to   explain why they  have  escaped detection so
far, their allowed    size is however    highly  model-dependent.  For
example,  if gravity is   the only force  that  feels the existence of
additional space dimensions,  the size of  the compactification radius
$R$ could be as large   as $10^{-3}$~mm~\cite{ADD}, without being   in
conflict       with  phenomenological      limits     from    collider
experiments~\cite{GRW}  and  cosmological constraints~\cite{HS}.  This
bound gets much  stronger, if fields  charged under the Standard Model
(SM) gauge group propagate  in the extra  dimensions as  well.  Again,
the actual value of the lower bound on the compactification scale $M =
R^{-1}$ crucially depends  on  the details of the  model.   If  all SM
fields experience  the existence of  one extra compact dimension as in
the          so-called           universal           extra-dimensional
scenario~\cite{IA,DDG,ACD,UED,OPS}, the lower limit  on $M$ is  rather
weak.   Mainly from   the    KK loop  contributions   to  the   $\rho$
parameter~\cite{ACD} and   the  decay $Z\to b\bar{b}$~\cite{OPS},  one
finds  $M \stackrel{>}{{}_\sim} 300$~GeV.  However,  if some of the SM
fields,  in particular  the   fermions, are confined to   the familiar
4-dimensional world, $M$ is constrained  much more strongly, namely $M
\stackrel{>}{{}_\sim}     4$~TeV~\cite{MPR}.  This order of  magnitude
increase of the bound  in non-universal 5-dimensional~(5D) settings of
the SM can be attributed to the fact that single KK excitations couple
at  tree-level to light  SM modes  on  the brane. In universal  models
these couplings are forbidden by selection rules.

In this paper we improve the constraints on the compactification scale
$M$ derived earlier~\cite{MPR} in   a wide class of  non-universal  5D
models by  taking the latest  LEP2 data into account.  Furthermore, we
carry out a multi-parameter global  fit of $M$ simultaneously with the
SM  parameters.  This  allows us to  properly include the correlations
of  the  compactification scale with  the SM parameters, in particular
with  the  Higgs-boson  mass $m_H$   and  the  top-quark mass   $m_t$.
In~\cite{RW}, it has been argued that the value of $m_H$, bounded from
below by direct Higgs  searches ($m_{H_{\rm SM}} \stackrel{>}{{}_\sim}
114$~GeV)~\cite{PDG2002}   and from  above  by perturbative  unitarity
($m_{H_{\rm   SM}} \stackrel{<}{{}_\sim} 1$~TeV)~\cite{LQT}, may  have
significant effects on the limits on $M$  derived by a global analysis
of electroweak precision    data,  and  vice   versa.    Finally,   we
systematically investigate the sensitivity of  future experiments at a
500-800~GeV linear $e^+e^-$ collider  such as TESLA. In this analysis,
we also  study   the improvements   which  can be  expected  from  the
so-called GigaZ option of TESLA, where the machine  is operated at the
$Z$ pole with a luminosity 100 times larger than that of LEP.

The general theoretical framework of our investigations is provided by
5D extensions of the SM  (5DSM) compactified on an $S^1/Z_2$ orbifold,
where all fermions are  localized  on one  of  the two orbifold  fixed
points.  As far  as the gauge fields are  concerned, we consider three
possible  scenarios:   (i) all gauge    fields propagate in  the extra
dimension, i.e.,  the bulk; (ii) only the  SU(2)$_L$  gauge bosons are
bulk fields, while  the  U(1)$_Y$  gauge  field  is confined to    the
orbifold fixed point where the fermions  live; (iii) only the U(1)$_Y$
boson propagates in     the  bulk, while  the   SU(2)$_L$   bosons are
restricted  to the brane.  As  has been shown in~\cite{MPR}, the above
5D models     can  be consistently  quantized    using  appropriate 5D
gauge-fixing conditions  that  lead, after the  KK  reduction,  to the
known class of $R_\xi$ gauges.

Although   it  is not  our    intention  to  put  forward an  explicit
string-theoretic construction  of  these non-universal 5D   models, we
note that they may result from  the intersection of higher-dimensional
$p$-branes~\cite{JP,BDL}  within the context  of  type  I and type  II
string  theories~\cite{Dmodels}.  The   SU(2)$_L$  and U(1)$_Y$  gauge
groups of the SM   may be associated  with two  separate  intersecting
higher-dimensional    spaces   ($Dp$  branes)     that  have different
compactification radii.  If the  SU(2)$_L$ compactification radius  is
so small that the KK  states become heavy  enough to decouple from the
low-energy observables of interest,   the low-energy sector of such  a
model would  effectively look like  a   scenario with SU(2)$_L$  gauge
bosons confined  to a 3-dimensional  brane.  In this construction, all
SM fermions are assumed  to be  localized on  the intersection of  the
SU(2)$_L$   and U(1)$_Y$     branes that   constitutes our  observable
3-dimensional world.  Localized brane interactions may also be induced
by radiative effects from KK bosons in the bulk.   They are omitted in
our  analysis.  Detailed discussions   on   this topic  may  be  found
in~\cite{GGH}.

The article  is  organized as follows.   In Section~\ref{Framework} we
summarize the main features of the minimal non-universal 5D extensions
of    the  SM  studied   in detail    in~\cite{MPR},  and  sketch  the
phenomenological   consequences.      Section~\ref{LEP1} describes our
approach  to  probing 5D   models and presents  the constraints on the   
compactification scale  $M$ obtained from  multi-parameter fits     to  
electroweak   precision   data,      including   error   correlations.  
In Section~\ref{LEP2}   we   analyze   the  5D     effects   on    the
total cross sections and asymmetries  of  fermion-pair  and   $W$-pair
production  at LEP2 and evaluate the   corresponding bounds on $M$. In
addition,  we  give   combined   electroweak  and  LEP2  bounds.   The
sensitivity to $M$ at a  future $e^+e^-$ linear collider is  discussed
in   Section~\ref{NLC}.  Section~\ref{Conclusions} highlights the main
conclusions. Finally, the  novel couplings of the  KK  bosons entering
$W$-pair production  and Higgsstrahlung are  presented in Appendices A
and B.

\setcounter{equation}{0}
\section{Theoretical Framework}\label{Framework}

In this section, we briefly review the minimal 5D extensions of the SM
under study.  Further details may be found in~\cite{MPR}.  We start by
considering the   so-called bulk-bulk  model where  all   gauge fields
propagate in the extra dimension.  The gauge and  Higgs sector of this
model is described by the 5D Lagrangian
\begin{eqnarray}
\label{bulkbulkLagrangian}
\la(x,y) &=&  - \, \frac{1}{4} \, B_{M N} \, B^{M N}\: 
              - \: \frac{1}{4} \, F^a_{M N} F^{a M N} \:
	      + \: \big( D_M \, \Phi_1\big)^\dagger\, 
              \big(D^M\,\Phi_1\big)\nonumber\\  
         &&  + \: \delta(y) \big( D_\mu \, \Phi_2\big)^\dagger \, 
	           \big(D^\mu\,\Phi_2\big)\, , 
\end{eqnarray}
where $B_{M  N}$ denotes the  U(1)$_Y$ field  strength and $F^a_{M N}$
($a = 1,2,3$) the SU(2)$_L$ field strength.  The covariant  derivative
$D_M$ is defined by
\begin{equation}
\label{covariantderivative}
D_M\ = \ \partial_M \, - \, i \, \frac{g_5}{2} \, A^a_M \, \tau^a \, - i \,
\frac{g_5'}{2} \, B_M\, , 
\end{equation}
where $\tau^a$ denote    the  Pauli  matrices, and    analogously  for
$D_{\mu}$.    In   (\ref{bulkbulkLagrangian})  the   Higgs  potential,
gauge-fixing and ghost terms  are omitted for brevity.  Throughout the
paper, the 5D~Lorentz indices are denoted  with capital Roman letters,
in the above~$M,N =   0,1,2,3,5$, while for the  respective 4D~indices
Greek  letters  are  used, in  the above~$\mu  = 0,1,2,3$.  The  Higgs
doublet $\Phi_2$ is restricted to  a brane at the orbifold fixed-point
$y=0$,   while the doublet   $\Phi_1$  propagates in   the bulk.   The
zero-mode of   $\Phi_1$  and $\Phi_2$ acquire  the  vacuum expectation
values (VEV) $v_1$ and $v_2$, respectively.  As usual, we define $\tan
\beta =   v_2/v_1$   and   $v =    \sqrt{v_1^2 +   v_2^2}$.   In   the
phenomenological analysis we will often focus on the cases $\sin \beta
= 0$ or $\sin \beta = 1$.

Different  minimal  5D   extensions of the  SM    can be  obtained  by
restricting either the SU(2)$_L$  or the U(1)$_Y$  gauge boson to  the
brane at $y = 0$. In the first case, one has
\begin{equation}
\label{branebulkLagrangian}
\la(x,y)   =  - \, \frac{1}{4} \, B_{M N} \, B^{M N} \ 
              + \, \delta(y) \lreckig{ 
	      - \, \frac{1}{4} \, F^a_{\mu \nu} F^{a \mu \nu} \ 
              + \, \lr{D_{\mu} \, \Phi}\da \, \lr{D^{\mu} \, \Phi} }\, ,
\end{equation}
while in the second model
\begin{equation}
\label{bulkbraneLagrangian}
\la(x,y)   =  - \, \frac{1}{4} \, F^a_{M N} F^{a M N} \ 
              + \, \delta(y) \lreckig{ 
	      - \, \frac{1}{4} \, B_{\mu \nu} \, B^{\mu \nu} \ 
              + \, \lr{D_{\mu} \, \Phi}\da \, \lr{D^{\mu} \, \Phi} }\, .
\end{equation}
We will refer  to (\ref{branebulkLagrangian}) as the  brane-bulk model
and to (\ref{bulkbraneLagrangian})  as the bulk-brane model. Here, any
Higgs  doublet  has  to  be  confined  to the brane  because  of gauge 
invariance.   In   all  the   above  models,  the fermionic degrees of 
freedom are localized on the brane.

Compactification and integration over the extra dimension is performed
most easily by expanding the bulk fields in KK modes which respect the
symmetries  of the $S^1/Z_2$ orbifold.   The $Z_2$-parity  of the bulk
fields is chosen  such that   the  theory is  gauge invariant  at  the
classical level and the light degrees of freedom  (the zero modes of a
Fourier expansion) coincide with those of the SM spectrum.

The  resulting effective 4-dimensional theory contains heavy KK vector
and  scalar  particles   with   masses  that  are  multiples   of  the 
compactification scale $M$, in addition to the usual SM   degrees   of 
freedom. Electroweak symmetry breaking by a VEV of a brane Higgs field
leads to mixing between different KK-modes including   the  zero modes
which shifts  the masses and  the couplings  to fermions. These shifts
are especially important for observables at the $Z$ pole where effects
from  the exchange of heavier KK  modes,  dominating at high energies,
are negligible.

In  our  phenomenological  analysis,   we  proceed  as  follows.   The 
prediction   of    the     5DSM     for     a     given     observable  
$\mathcal{O}^{\mr{5DSM}}$    is    related   to   the   SM  prediction 
$\mathcal{O}^{\mr{SM}}$ by
\begin{equation}
\label{generalformofpredictions}
{\cal O}^{\rm 5DSM} \ =\ {\cal O}^{\rm SM} \, 
                    \big( 1\: +\: \Delta^{\rm 5DSM}_{\cal O} \big)\, ,
\end{equation}
where $\Delta^{\rm 5DSM}_{\cal O}$ is the tree-level effect due to the
compactified extra    dimension.  The SM  radiative   corrections  are
included in $\mathcal{O}^{\mr{SM}}$.   However,   SM loop  effects  on
$\Delta^{\rm 5DSM}_{\cal O}$ as well as KK loop effects are neglected.
For   compactification  scales in      the TeV  range  this    is well
justified. The  tree-level calculation of $\Delta^{\rm 5DSM}_{\cal O}$
is performed in terms of the compactification scale  $M$ and the usual
SM   input parameters, that   is the   electromagnetic  fine structure
constant~$\alpha$, the Fermi constant   $G_F$, and the $Z$-boson  mass
$m_{Z (0)}$. The index $(0)$ indicates that  the observed $Z$ boson is
to be identified with the lightest mode of the corresponding KK tower.
The  remaining  SM  parameters  $m_t$, $m_H$,  and the strong coupling 
constant $\alpha_s(m_Z)$ enter in $\mathcal{O}^{\mr{SM}}$, but do  not 
influence  the  calculation  of  $\Delta^{\rm 5DSM}_{\cal O}$  in this 
approximation.

Furthermore,  one  has to  take  into account  that the tree relations
between the SM  input  parameters and  other SM  parameters like gauge
couplings  and  VEVs are   also affected by  the  extra  dimension. An
exception  is  the fine structure constant  
\begin{equation}
\label{alphadef}
\alpha   = e^2/4 \pi \, . 
\end{equation}
To order $1/M^2$, one finds
\begin{equation}
\label{deltazdef}
m^2_{Z (0)} \  =\ \frac{(g^2 +g'{}^2)v^2}{4} \, \lr{1 \: + \: \Delta_{Z}\,X}\, 
\end{equation}
and
\begin{equation}
\label{deltagfermidef}
G_F \ = \ \frac{\pi \alpha}{\sqrt{2} \sin^2 \theta_W \, 
       \cos^2 \theta_W \, m_{Z (0)}^2} \, \lr{ 1\: +\: \Delta_G\, X}\,,
\end{equation}
where  
\begin{equation}
\label{gaugecoupl}
g = g_5/\sqrt{2 \pi R} = e/\sin \theta_W \, , \quad 
g' = g'_5/\sqrt{2 \pi R} = e/\cos \theta_W \, 
\end{equation}
are the usual 4D gauge couplings. Inverting (\ref{deltagfermidef}), one obtains
\begin{equation}
\label{thetawhatthroughthetaw}
\sin^2 \theta_W \ = \ \sin^2 \hat{\theta}_W \, \lr{ 1\: -\: 
\frac{\cos^2 \hat{\theta}_W}{\sin^2 \hat{\theta}_W - \cos^2 \hat{\theta}_W} 
\, \Delta_{G} \,X} \, ,
\end{equation}
where $\hat{\theta}_W$ denotes the SM value of the weak mixing angle
following from 
\begin{equation}
\label{fermidefsm}
G_F \ = \ \frac{\pi \alpha}{\sqrt{2} \sin^2 \hat{\theta}_W \, 
       \cos^2 \hat{\theta}_W \, m_{Z (0)}^2} \,.
\end{equation}
In the above, the  shifts  with respect  to the  SM
relations    are    parameterized    by   the    model-dependent,  but
mass-independent   coefficients  $\Delta_Z$ and $\Delta_G$
and the dimensionless factor
\begin{equation}
\label{Xdefinition}
X= \frac{\pi^2}{3} \, \frac{m_{Z (0)}^2}{M^2} \ll 1 \, .
\end{equation}
For the bulk-bulk, the brane-bulk,  and the bulk-brane model  one has,
respectively,
\begin{eqnarray}
\label{deltaZ}
\Delta_{Z} &=& \bigg\{  - \, s^4_\beta\,,\ 
                        - \, \hat{s}^2_W\,,\ 
		        - \, \hat{c}^2_W\, 
	       \bigg\}\,, \\
\label{deltaG}
\Delta_{G} &=& \bigg\{ \, \hat{c}^2_W \bigg( 1\: -\: 2 s^2_{\beta}\: -\:
              \frac{\hat{s}^2_W}{\hat{c}^2_W} s^4_{\beta}\,\bigg)\, ,\  
              -\,\hat{s}^2_W\,,\  -\,\hat{c}^2_W \, \bigg\} \, ,
\end{eqnarray} 
where the abbreviations of  the  trigonometric functions are obvious.
It is the shift in $\sin^2 \theta_W$ given in (\ref{thetawhatthroughthetaw}) 
which provides very sensitive 
tests of the above 5D~models as will be seen in Section~\ref{GigaZ}. 
Similarly for other observables, we   will also expand 
the shift $\Delta^{\rm  5DSM}_{\cal O}$ in $X$  and keep only  the linear
term.   Exact  analytic  expressions  for 5D    shifts in  masses  and
couplings to all orders in $X$ are presented in~\cite{MPR}.

After compactification,  the couplings of   the KK modes of the  gauge
bosons to the SM  brane fermions are  determined  by their  SM quantum
numbers.   In   the interaction or weak    basis,  these couplings are
generically given by
\begin{equation}
\label{interactionsfermwithgbmodesabelianmodel}
\la_{\rm int}(x)\ = \ g \, \Psibar\, \gamma^{\mu} \, \lr{g_V
                        - g_A \gamma^5} \, \Psi\, \Big( A_{(0)
                        \mu} \, + \, \sqrt{2} \, \sum_{n=1}^{\infty}
                        A_{(n) \mu} \Big) \, ,
\end{equation}
where $g_V$ and $g_A$
are the usual vector and axial vector  coupling constants, and $A_{(n)
\mu}$  denotes the  {\it    {\rm n}th}  KK mode    of  a given   gauge
field. However, in  the presence  of a  nonzero VEV  of a brane  Higgs
field,   the  KK  Fourier-modes   mix     to form mass     eigenstates
$\hat{A}_{(n) \mu}$.  The couplings of these physical   fields can then be
parameterized as follows:
\begin{equation}
\label{masseigenstatelagrangian}
\la_{\rm int}(x)\ = \ \sum_{n=0}^{\infty} g_{(n)} \, \Psibar\, 
                        \gamma^{\mu} \, \lr{g_{V (n)}
                        - g_{A (n)} \gamma^5} \, \Psi\, 
			\hat{A}_{(n) \mu} \,  \, .
\end{equation}
For $\hat{A}_{(n) \mu} = W_{(n) \mu}$, one  has $g_{V (n)} = g_{A (n)}
= 1$ and $g_{(n)}  = g_{W (n)}/(2 \sqrt{2})$,  where for  the observed
$W$~boson
\begin{equation}
g_{W (0)} \ = \ g \, \bigg\{ 1\: -\: s^2_{\beta} \, \hat{c}^2_W\, X \, , 
                       \quad 1 \, , \quad  1\: -\: \hat{c}^2_W\, X \bigg\}
\end{equation}
in the bulk-bulk, brane-bulk, and  bulk-brane model, respectively.  In
the brane-bulk model,  there are of course  no higher KK  modes of the
$W$.  Similarly for the $Z$ modes, one obtains $g_{V (n)} = T_{3f (n)}
- 2 Q_{f (n)} s_W^2$,            $g_{A (n)} = T_{3f (n)}$,         and 
$g_{(n)} = g_{Z (n)}/2 c_W$, where, focusing on the observed $Z$ boson,
\begin{equation}
\label{zeromodecouplings}
\begin{split}
g_{Z(0)} \, & = \, \, \, g \, \, \, \bigg\{ 1\: -\: s^2_{\beta} \, X \, , 
                \quad 1 \, , \quad 1\bigg\} \, , \\
T_{3f (0)} \, & = \, T_{3f} \bigg\{ 1 \, , \quad  1 \, - \, \hat{s}^2_W \, X \, , 
                  \quad 1 \, - \, \hat{c}^2_W \, X \bigg\} \, , \\
Q_{f (0)} \, & = \, Q_f \, \bigg\{ 1 \, , \quad 1 \, - \, X \, , 
                 \quad 1 \bigg\} \, .
\end{split}
\end{equation}
Here, $Q_f$   and $T_{3f}$ are the  fermion  charge  ($Q_e =  -1$) and
3-component of the weak isospin ($T_{3 e} = -1/2$).  The photon is not
affected by electroweak symmetry breaking and, hence, $g_{V (n)} = 1$,
$g_{A (n)} = 0$, and $e_{(n)} = e$ for $n = 0$ and $\sqrt{2} e$ for $n
\ge 1$. In the brane-bulk and the bulk-brane models, the higher photon
KK modes are absent. The  coupling parameters $g_{W(n)}$,  $g_{Z(n)}$,
$T_{3f(n)}$, and $Q_{f(n)}$ for the higher KK modes  ($n \ge 1$)   can 
also be obtained from Appendix B of~\cite{MPR}. To first order in $X$, 
they are given by
\begin{equation}
\label{highermodecouplings}
\begin{split}
g_{W (n)} \, &  = \, \,  g \, \, \, \bigg\{ 
\sqrt{2} \left( 1\: -\: \frac{3}{2 \pi^2 n^2} \, s^2_{\beta} \, \hat{c}^2_W \, X \right), \quad 
\mathrm{no \, \, KK \, \, modes} \, , \quad 
\sqrt{2} \left( 1 \: -\: \frac{3}{2 \pi^2 n^2} \, \hat{c}^2_W \, X \right) \bigg\} \, , \\[1ex]
g_{Z(n)} \, & = \, \, \, g \, \, \, \bigg\{ 
\sqrt{2} \left( 1\: -\: \frac{3}{2 \pi^2 n^2} \, s^2_{\beta} \, X \right), \quad 
1 \, , \quad 
1 \bigg\} \, , \\[1ex]
T_{3f (n)} \, & = \, T_{3f}         \bigg\{ 
1 , \, \,  
\sqrt{2} \, s_W \left( 1 + 
       (\hat{c}^2_W - \frac{1}{2} \hat{s}_W^2) \frac{3}{\pi^2 n^2} \, X \right) , \, \, 
\sqrt{2} \, c_W \left( 1 + 
       (\hat{s}^2_W - \frac{1}{2} \hat{c}_W^2) \frac{3}{\pi^2 n^2} \, X \right) \bigg\} , \\[1ex]
Q_{f (n)} \, & = \, Q_f \,          \bigg\{ 
1 , \quad 
\frac{\sqrt{2}}{s_W} \left( 1\: - \: \hat{s}^2_W \frac{3}{2 \pi^2 n^2} \, X \right)\, , \quad 
\sqrt{2} \, \hat{c}_W \, \frac{3}{\pi^2 n^2} \, X \bigg\} \, .
\end{split}
\end{equation}

\setcounter{equation}{0}
\section{Electroweak Constraints Revisited}\label{LEP1}

In order to extract bounds  on $M$ from the   available data, one  can
proceed in two  ways. For the set  of precision  observables specified
in~\cite{MPR}  (mostly from LEP),  one may  fix  the SM parameters  at
their  current best  fit  values  and  calculate the 5DSM  predictions
${\cal    O}^{\rm 5DSM}$  using  (\ref{generalformofpredictions})  and
taking       the     SM        expectations    $\mathcal{O}^{\mr{SM}}$      
from~\cite{PDG2002}.   Then,   one   can    perform   a  one-parameter
$\chi^2$-analysis  for   $M$   or  equivalently   $X$     defined   in
(\ref{Xdefinition}). The $\chi^2$-function is given by
\begin{equation} 
\label{chisquare}
\chi^2 (X) = \sum_{i,j} \, 
             \lr{ \mathcal{O}_i^{\mr{exp}}\: -\: \mathcal{O}_i^{\mr{5DSM}} } 
             V^{-1}_{ij} 
	     \lr{ \mathcal{O}_j^{\mr{exp}}\: -\: \mathcal{O}_j^{\mr{5DSM}} }
\end{equation}
with the covariance matrix $V_{ij} = \Delta \mathcal{O}_i \, \rho_{ij}
\, \Delta \mathcal{O}_j$, $\Delta \mathcal{O}_i$ being the measurement
error of a given observable $\mathcal{O}_i^{\mr{exp}}$ and $\rho_{ij}$
being the matrix of  correlation coefficients. This approach is widely
used in  the literature. It  has also  been followed in  our  previous
analysis~\cite{MPR}.  However, in this approach  possible correlations
between  the  SM  parameters  and  the size of the extra dimension are 
ignored, and hence the  bounds on $M$ may be overestimated. Therefore,
it  is interesting to follow  a more general approach in  which $X$ is 
fitted        simultaneously     with      the      SM      parameters  
$\alpha_{\mr{em}}(m_Z)$, $G_F$, $m_Z$, $\alpha_s(m_Z)$, $m_t$,   $m_H$ 
to the data. The SM  predictions     $\mathcal{O}_i^{\mr{SM}}$      in
(\ref{generalformofpredictions})       are       obtained         from
ZFITTER~\cite{ZFITTER},  while    the 5D    corrections   $\Delta^{\rm
5DSM}_{\cal  O}$   are   taken from~\cite{MPR}.   The  multi-parameter
minimization    of  $\chi^2$     is   performed   with    the  program
MINUIT~\cite{MINUIT}.

The bounds   on $X$ can be   derived from (\ref{chisquare})  by either
using   Bayesian   statistics\footnote{ For   example,    in  Bayesian
statistics with a flat prior  in the physical region  $X \ge 0$ and  a
zero  prior       in     the     unphysical region,        the    
95\%~(1.96~$\sigma$)~c.l.  bound   $X_{95}$  is   given by  
$0.95 = \int^{X_{95}}_0 \, d X \, \, P(X)/
\int^{\infty}_0 \, d X \, \, P(X) \, ,$ 
where $P(X) = \exp [- (\chi^2(X) - \chi^2_{\mr{min}})/2]$.  } or by
simply requiring
\begin{equation}
\label{deltachibound}
\Delta \chi^2 = \chi^2  (X) -  \chi^2_{\mr{min}} < n^2 \, 
\end{equation}
for $X$ not to be  excluded  at  the  $n \, \sigma$  confidence level.
In the above, $\chi^2 (X)$ is the minimum for a given $X$ with respect
to  all other fit   parameters  requiring  $m_H  \ge 114$~GeV,   while
$\chi^2_{\mr{min}}$  is the overall  minimum in the physically allowed
region $X \ge 0$, $m_H \ge  114$~GeV. If the best  fit value of $X$ is
not  too far in  the unphysical region,  both  methods lead to similar
results and   approximate      well the  results   of   the    unified
approach~\cite{FC}.    We  will present  the  bounds  as obtained from
(\ref{deltachibound}).

\begin{figure}[t]
\mbox{} \hspace{-11pt} \includegraphics[width=16.4cm]{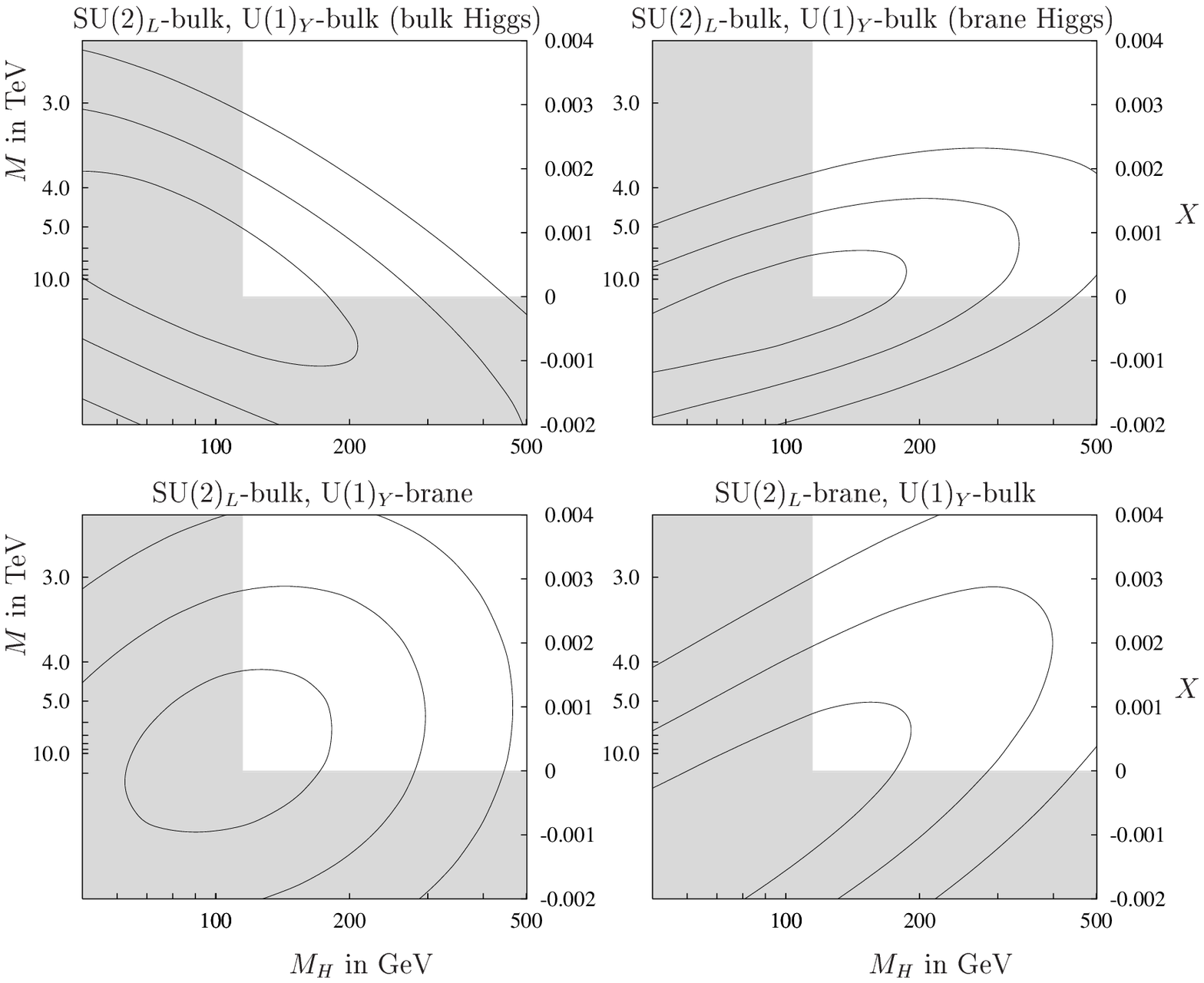}
\mycaption{\label{contoursprecobs} Contours of 
$\Delta \chi^2  = 1,\, 4,\, 9$ (eq.   \ref{deltachibound}) derived
from  multi-parameter  fits  to electroweak precision data. The shaded
regions of the parameter space  correspond  to $m_H < 114$~GeV  and/or
$M^2 < 0$.}
\end{figure}

\begin{table}[t] 
\mbox{}
\vspace{-1cm}
\begin{center} 
\begin{tabular}{c|c|c|c|c} 
\hline\hline 
&  \pb{2.4}{-0.05}{ \vspace{-0.2cm}  brane-bulk } &  
   \pb{2.4}{-0.05}{ \vspace{-0.2cm}  bulk-brane } &  
   \pb{2.5}{-0.05}{ \vspace{-0.2cm}  bulk-bulk \\ (brane  Higgs) } &  
   \pb{2.5}{-0.05}{ \vspace{-0.2cm}  bulk-bulk \\ (bulk Higgs) } 
   \\ \hline \hline
   \pb{3.5}{-0.}{one-parameter fit} & 
   \pb{2}{-0.}{4.9} & 
   \pb{2}{0.}{3.2} & 
   \pb{2}{-0.}{5.5} & 
   \pb{2}{0.}{4.2} \\ \hline
   \pb{3.5}{-0.}{multi-parameter fit} & 
   \pb{2}{-0.}{3.1} & 
   \pb{2}{0.}{3.1} & 
   \pb{2}{-0.}{4.2} & 
   \pb{2}{0.}{3.7} \\ \hline \hline
\end{tabular} 
\end{center} 
\mycaption{\label{correlations} 2$\sigma$ bounds on $M$ in TeV 
derived from electroweak precision measurements~\cite{PDG2002}.} 
\end{table} 

Having described the methods, we will now discuss the resulting bounds
on the compactification  scale $M$ listed in Table \ref{correlations}.
Superseding  the  one-parameter  fit in~\cite{MPR},  we use the latest
experimental  data~\cite{PDG2002}   and  take into account   the error
correlations  which     had  previously   been ignored.    The   error
correlations have only a small effect, shifting  the bounds by no more
than 0.2~TeV.   However, the  change in the  data  with respect to the
data of  the year 2000  alters the bounds  by as much  as 1~TeV.  This
mainly is due  to the large  shift of  the experimental  value for the
forward-backward asymmetry $A^{(0,b)}_{FB}$  which is now found to  be
more than   3$\sigma$ below the SM  expectation.   Only  the bulk-bulk
model with  a bulk Higgs  predicts a smaller value of $A^{(0,b)}_{FB}$
than the SM.  Correspondingly,  the bound  for  this model  is lowered
while  the constraints  on the  other  models become stronger. 

In the multi-parameter fit, the correlations between $M$ and    the SM
parameters  reduce  the bounds  as expected, the   size  of the effect
varying  from model to  model. In  the brane-bulk  model the effect is
biggest lowering the bound by almost 40\%. Here, the best fit value is
relatively far in the unphysical  region. Thus,  as a cross-check,  we
also performed  a Bayesian analysis leading  to   a  bound   which  is 
0.1~TeV below the bound from (\ref{deltachibound}).

Most interesting is the correlation between the compactification scale
and the mass  of the Higgs boson. This  correlation is  illustrated in
Fig.~\ref{contoursprecobs}.  The  data  set  used  for  this  analysis 
differs from the one used for the familiar blue-band plot~\cite{EWWG}. 
Within the SM ($X=0$ in Fig.~\ref{contoursprecobs}) it leads to a best
fit  value $m_H = 100^{+70}_{-40}$~GeV   and  to the  2$\sigma$  upper
bound  $m_H < 280$~GeV.  As can be seen in Fig.~\ref{contoursprecobs}, 
in  all  models  with  the  Higgs  field  localized   on  the   brane,
the  existence  of  an  extra  dimension   favors   a   heavier  Higgs 
boson.    This  confirms  the  observation  for  the  bulk-bulk  model 
in~\cite{RW}.   The  effect  is  most  pronounced  in  the   bulk-bulk
model with a brane Higgs and in the brane-bulk  model. Quantitatively,
for $M = 5$~TeV,   the   best   fit  values  increase   to      $m_H = 
170^{+105}_{-60}$~GeV and $m_H = 155^{+105}_{-60}$~GeV,  respectively.   
If the compactification scale  is  included  in   the  multi-parameter
fit, the 2$\sigma$ upper bound on $m_H$ is relaxed to 330 and 400~GeV, 
respectively.

\setcounter{equation}{0}
\section{LEP2 Constraints}\label{LEP2}

\begin{figure}[t]
\begin{center} 
\includegraphics[width=12cm]{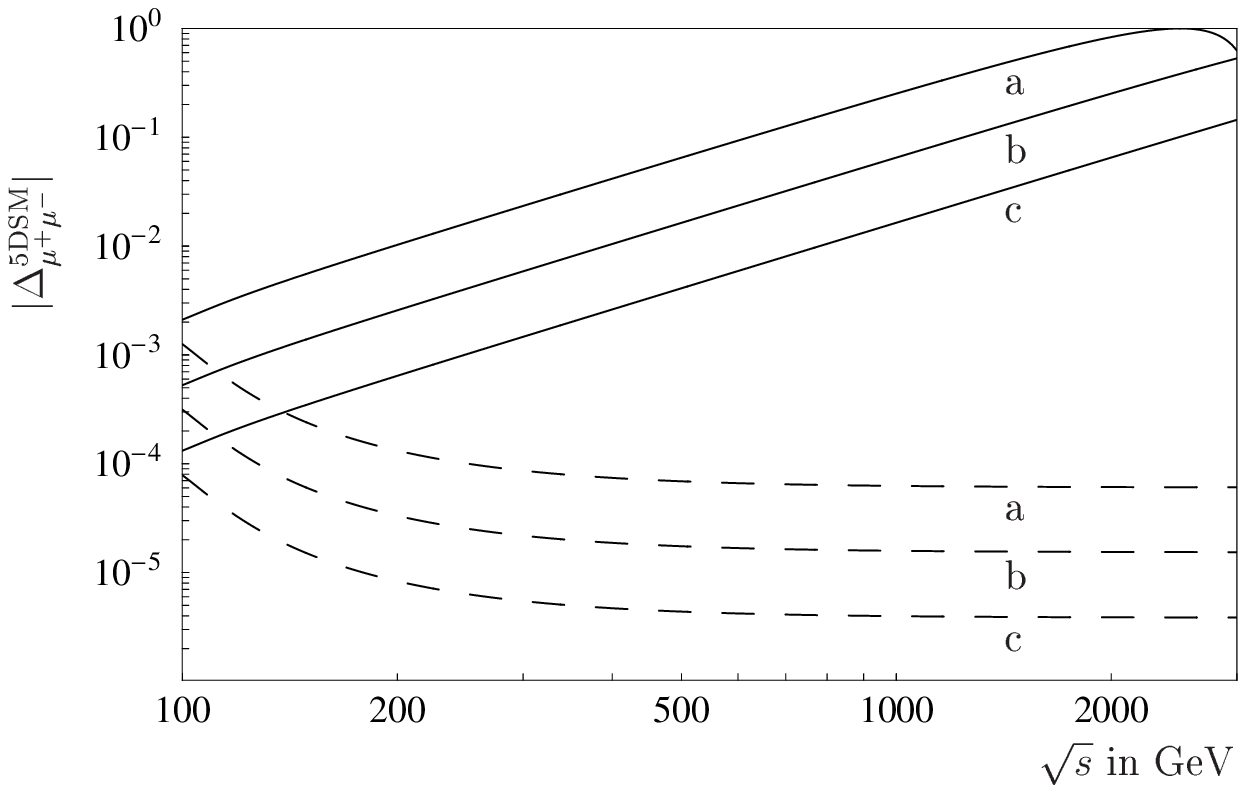}
\end{center}
\mycaption{\label{sdependence} The shift $\Delta^{\mr{5DSM}}_{\mu^+ \mu^-}$ 
of the total cross section  for muon-pair production in  the bulk-bulk
model with a brane Higgs  for $M  =   5$~(a), 10~(b), and  20~(c)~TeV.
The  dashed  curves  show  the  effects from the mixing  of masses and
couplings only.}
\end{figure}

At energies above the $Z$ pole, the virtual exchange of KK excitations
of  the  SM  gauge  bosons  becomes  dominant. As long as $s \ll M^2$,
the  main  effects  come  from  the   interference    of   zero   with
higher KK  modes. These effects scale  like $s/M^2$ in contrast to the
energy-independent  modifications  of   masses  and couplings,   as is
illustrated  in  Fig.~\ref{sdependence} for muon-pair  production. The
residual energy dependence of  the impact of  the latter  on the cross
section (dashed  curves  in    Fig.~\ref{sdependence}) is   due to the  
transition from dominant $Z$ exchange to dominant photon exchange.

Focusing  on LEP2, we have investigated  the  total cross sections for
lepton-pair     production,   hadron     production,     and    Bhabha
scattering. Forward-backward  asymmetries for muon and  tau production
as  well   as     the  heavy quark     observables   $A^{(0,b)}_{FB}$,
$A^{(0,c)}_{FB}$, $R_b =  \sigma(b  \overline{b}) / \sigma(\mr{had})$,
and $R_c = \sigma(c \overline{c}) /  \sigma(\mr{had})$ are included in
the fits, although they do  not contribute  noticeably to the  bounds.
For completeness, we have also investigated $W^+W^-$ production.
\subsection{Fermion-Pair Production}\label{FPP}

The differential cross section for fermion-pair production is given by
\begin{equation}
\label{diffcrosssection}
\begin{split}
\frac{d \, \sigma (e^+ e^- \to f \, \overline{f})}{d \, \cos \vartheta} = 
\frac{N_f s}{128 \, \pi} 
& \left[( 1 + \cos \vartheta)^2 \, (|M^{ef}_{LL} (s)|^2 + |M^{ef}_{RR} (s)|^2) 
                                                           \, \, + \right. \\
& \, \, \, \left. ( 1 - \cos \vartheta)^2 \, 
                                   (|M^{ef}_{LR} (s)|^2 + |M^{ef}_{RL} (s)|^2) 
                                                                  \right] \, ,
\end{split}
\end{equation}
where $\vartheta$ is    the  scattering angle between  the    incoming
electron and the negatively  charged outgoing fermion, and  $N_f=1(3)$
for   leptons (quarks) in  the  final state. For   the matrix elements
entering (\ref{diffcrosssection}) one finds
\begin{equation}
\label{effprop}
M^{ef}_{\alpha \beta} (s) =  
\sum_{n=0}^{\infty}  
\lr{ e_{(n)}^2 \, \frac{Q_e Q_f}{s - m^2_{\gamma (n)}} \, + \, 
\frac{g^e_{\alpha (n)} g^f_{\beta (n)}}{\cos^2 \theta_W} \, 
\frac{1}{s - m^2_{Z (n)}} }
\end{equation}
with the couplings
\begin{equation}
\label{couplings}
g^f_{L,R (n)} = \frac{g_{Z (n)}}{2} \left( g_{V (n)} \pm g_{A (n)} \right) \, 
\end{equation}
derived        from                  (\ref{masseigenstatelagrangian}), 
(\ref{zeromodecouplings}),       and  (\ref{highermodecouplings}). The   
expression   (\ref{diffcrosssection}) has also been used in~\cite{CL}. 
However, the  mixing  effects  in the couplings and masses included in
(\ref{effprop}) and (\ref{couplings}) have been neglected there.

\begin{table}[t]
\begin{center}
\begin{tabular}{c|c|c|c|c} 
\hline\hline 
& \pb{2.4}{-0.05}{ \vspace{-0.2cm} brane-bulk }
& \pb{2.4}{-0.05}{ \vspace{-0.2cm} bulk-brane }
& \pb{2.5}{-0.05}{ \vspace{-0.2cm} bulk-bulk \\ (brane Higgs) }
& \pb{2.5}{-0.05}{ \vspace{-0.2cm} bulk-bulk \\ (bulk Higgs)  } 
\\ \hline \hline  
$\mu^+ \mu^-$ & \pb{2}{0.}{ 2.0 } & \pb{2}{0.}{ 1.5 } 
& \pb{2}{0.}{ 2.5 } & \pb{2}{0.}{ 2.5 } \\ \hline
$\tau^+ \tau^-$ & \pb{2}{0.}{ 2.0 } & \pb{2}{0.}{ 1.5 } 
& \pb{2}{0.}{ 2.5 } & \pb{2}{0.}{ 2.5 } \\ \hline 
hadrons & \pb{2}{0.}{ 2.6 } & \pb{2}{0.}{ 4.7 } 
& \pb{2}{0.}{ 5.4 } & \pb{2}{0.}{ 5.8 } \\ \hline 
$e^+ e^-$ & \pb{2}{0.}{ 3.0 } & \pb{2}{0.}{ 2.0 } 
& \pb{2}{0.}{ 3.6 } & \pb{2}{0.}{ 3.5 } \\ \hline 
$W^+ W^-$ & \pb{2}{0.}{ 1.1 } & \pb{2}{0.}{ 2.0 } 
& \pb{2}{0.}{ 2.1 } & \pb{2}{0.}{ 2.2 } \\ \hline 
\parbox{2.9cm}{\begin{center} LEP2 combined \end{center}} & 
\pb{2}{0.}{ 3.5 } & \pb{2}{0.}{ 4.6 } 
& \pb{2}{0.}{ 5.6 } & \pb{2}{0.}{ 5.9 } \\ \hline 
\parbox{2.9cm}{\begin{center} Electroweak \\ and LEP2 data \\combined \end{center}} & 
\pb{2}{0.}{ 5.4 } & \pb{2}{0.}{ 4.8 } 
& \pb{2}{0.}{ 6.9 } & \pb{2}{0.}{ 6.0 } \\ \hline \hline   
\end{tabular}
\end{center}
\mycaption{\label{fermiondata} 2$\sigma$ bounds on $M$ in TeV 
from  a one-parameter fit  derived from LEP2  data alone and from LEP2
and electroweak precision data combined.}
\end{table}

The data taken  by the   LEP experiments  for  muon, tau,  and  hadron
production is properly  combined in~\cite{EWWG}  for energies  between
$\sqrt{s} = 130 \, \mr{GeV}$ and $\sqrt{s} = 207 \, \mr{GeV}$.  In the
hadronic channel  it is very important  to take into account the large
correlations  between  the data   at  different energies. If  they are
ignored, the bounds on the compactification scale are overestimated by
as much as 3 TeV.  On the other hand, correlations in the muon and tau
channels are extremely small and have little effect.

The  bounds from  a  simple one-parameter  analysis are summarized  in
Table~\ref{fermiondata}.  In the  muon and tau channel, the bulk-brane
model  is  least  restricted   because  essentially only   left-handed
fermions interact with the KK modes.  The best fit  values turn out to
lie always in the physical region $X \ge 0$.
Hadron production puts more stringent bounds  on the 5D models because
of the  larger cross section.  In this  case, the  brane-bulk model is
least  restricted,  since the hadronic  cross section  is dominated by
left-handed quarks whose small  hypercharges suppress the interference
effects with the U(1)$_Y$ KK modes. In this  case, the best fit values
lie in the unphysical region $X < 0$.

The differential cross  section for Bhabha scattering may conveniently
be expressed as
\begin{equation}
\label{diffcrosssectionBhabha}
\begin{split}
\frac{d \, \sigma (e^+ e^- \to e^+ e^-)}{d \, \cos \vartheta} = &
\frac{s}{128 \, \pi} 
\left[( 1 + \cos \vartheta)^2 \, (|M^{ee}_{LL} (s) + M^{ee}_{LL} (t)|^2 + 
|M^{ee}_{RR} (s) + M^{ee}_{RR} (t)|^2) \, \, + \right. \\
& \left. ( 1 - \cos \vartheta)^2 \, (|M^{ee}_{LR} (s)|^2 + |M^{ee}_{RL} (s)|^2) + 
4 \, (|M^{ee}_{LR} (t)|^2 + |M^{ee}_{RL} (t)|^2) \right] ,
\end{split}
\end{equation}
where $t  = - s (1 -  \cos \vartheta)/2$ and $M^{ee}_{\alpha \beta} (s
\, \mathrm{or} \, t)$ can be read off  from (\ref{effprop}). The total
cross       section       is         calculated    by      integrating
(\ref{diffcrosssectionBhabha})  over $\vartheta$  in the  experimental
ranges.      Since  the  Bhabha         data  of   the      four   LEP
experiments~\cite{Bhabhadata} has  not   yet  been  combined, possible
correlations of the different experiments cannot be accounted for,  at 
least for the time being.

As can be seen  from Table~\ref{fermiondata}, the  bounds on  $M$ from
Bhabha scattering are approximately 1~TeV stronger than those from the
other   leptonic  channels. This is   due  to the   large Bhabha cross
section.  On the other  hand, the dominance  of the $t$-channel photon
exchange, which is not affected by the presence of an extra dimension,
reduces the sensitivity of  Bhabha  scattering with respect  to hadron
production in almost all 5D models.

\subsection{$W^+W^-$ Production}\label{WW}
\label{wpairprod}

The     differential     cross   section    for   $W^+W^-$  production
reads~\cite{Alles}
\begin{equation}
\label{diffcrosssectionWW} 
\begin{split} 
& \frac{d \, \sigma (e^+ e^- \to W^+ \, W^-)}{d \, \cos \vartheta} =  
  \frac{1}{32 \pi s} \, \beta \\[2ex] 
& \, \times \left\{ \beta^2 \lreckig{M^2_L(s) + M^2_R(s)} s^2  
  \lreckig{\frac{s}{m_{W (0)}^2} + \sin^2 \vartheta\,  
  \lr{ \frac{3}{4} - \frac{s}{4 m_{W (0)}^2} + \frac{s^2}{16 m_{W (0)}^4}}} \right.\\[2ex]  
& \, \quad +
  M^2_L(t) t^2 \lreckig{ \frac{s}{4 m_{W (0)}^2} + \beta^2 \sin^2 \vartheta  
  \lr{ \frac{s^2}{16 t^2} + \frac{s^2}{64 m_{W (0)}^4}} } \\[2ex] 
& \left. \, \quad + M_L(t) M_L(s) \, s t \, \lreckig{  \! 2 +
  2 \frac{m_{W (0)}^2}{t}  + \beta^2 \frac{s}{m_{W (0)}^2} - \beta^2 \sin^2 \vartheta
  \lr{ \frac{s}{4 t} + \frac{s}{8 m_{W (0)}^2} - \frac{s^2}{16 m_{W (0)}^4} } \!  }   
  \right\} \, ,         
\end{split}
\end{equation} 
where $\beta = \sqrt{1 - 4 m_{W (0)}^2/s}$, 
$t= m^2_{W (0)} - s \, ( 1 - \sqrt{1 - \beta} \, \cos \vartheta)/2$ 
and $\vartheta$ is the scattering
angle  between  the  electron and the    negatively charged $W$ boson.
Note, that $m_{W (0)} = m_W^{\mr{SM}}  \, \lr{1 + \Delta_{m_W}
X}$ with
\begin{equation}
\label{mwshift}
\Delta_{m_W} = 
\bigg\{ \ 
 \frac{1}{2}\, s^4_{\beta} \hat{s}^2_W \, - \,
 \frac{\hat{s}^2_W \hat{c}^2_W}{2 \hat{c}_{2  W}}\,
 \bigg(\,1\:  -\: 2 s^2_{\beta}\: -\: 
                     \frac{\hat{s}^2_W}{\hat{c}^2_W} s^4_{\beta}\,\bigg) \,,\quad 
 \frac{\hat{s}^2_W \hat{c}^2_W}{2\, \hat{c}_{2W}}\,,\quad
 \frac{\hat{s}^2_W \hat{c}^2_W}{2\, \hat{c}_{2W}}\, 
\bigg\}\, 
\end{equation} 
for the    bulk-bulk,   the brane-bulk,  and   the  bulk-brane  model,
respectively~\cite{MPR}.   Furthermore, $M_{L,R}(s)$  and $M_L(t)$ are
given by
\begin{equation}  
\begin{split}  
M_{\alpha}(s) & = \sum_{n=0}^{\infty} 
\lr{ \frac{Q_e \, e_{(n)} \, g_{3 (n)}^{\gamma}} {s - m^2_{\gamma (n)}} \, + 
  \, \frac{g^e_{\alpha (n)} g^{Z}_{3 (n)}}{\cos \theta_W} 
  \, \frac{1}{s - m^2_{Z (n)}} } \, , \\[2ex] 
M_{L}(t) & = \frac{g^2_{W (0)}}{t} \, ,
\end{split}  
\end{equation} 
while   $M_{R}(t)  = 0$ as  in the  SM. A novel feature are the triple
gauge-boson couplings $g_{3 (n)}^{\gamma}$ and $g^{Z}_{3  (n)}$ of the
photon, the $Z$ boson, and their respective KK modes with the $W$ zero
modes.    They   are  given  in   Appendix~\ref{wpair}  along with the 
corresponding Feynman rules.

As can be  seen  from Table~\ref{fermiondata}, $W$-pair  production at
LEP2 provides relatively weak bounds on $M$. This can be understood by
realizing that the effects of KK exchange are almost negligible due to
the   suppression of the  interference of   SM and KK   exchange by an
additional factor $X$. This is a direct consequence of selection rules
which forbid the triple boson couplings of a single  higher KK mode to
zero modes for the   gauge eigenstates such   that  they can only   be
induced by mixing.

\begin{figure}[t]
\mbox{} \hspace{-11pt} \includegraphics[width=16.4cm]{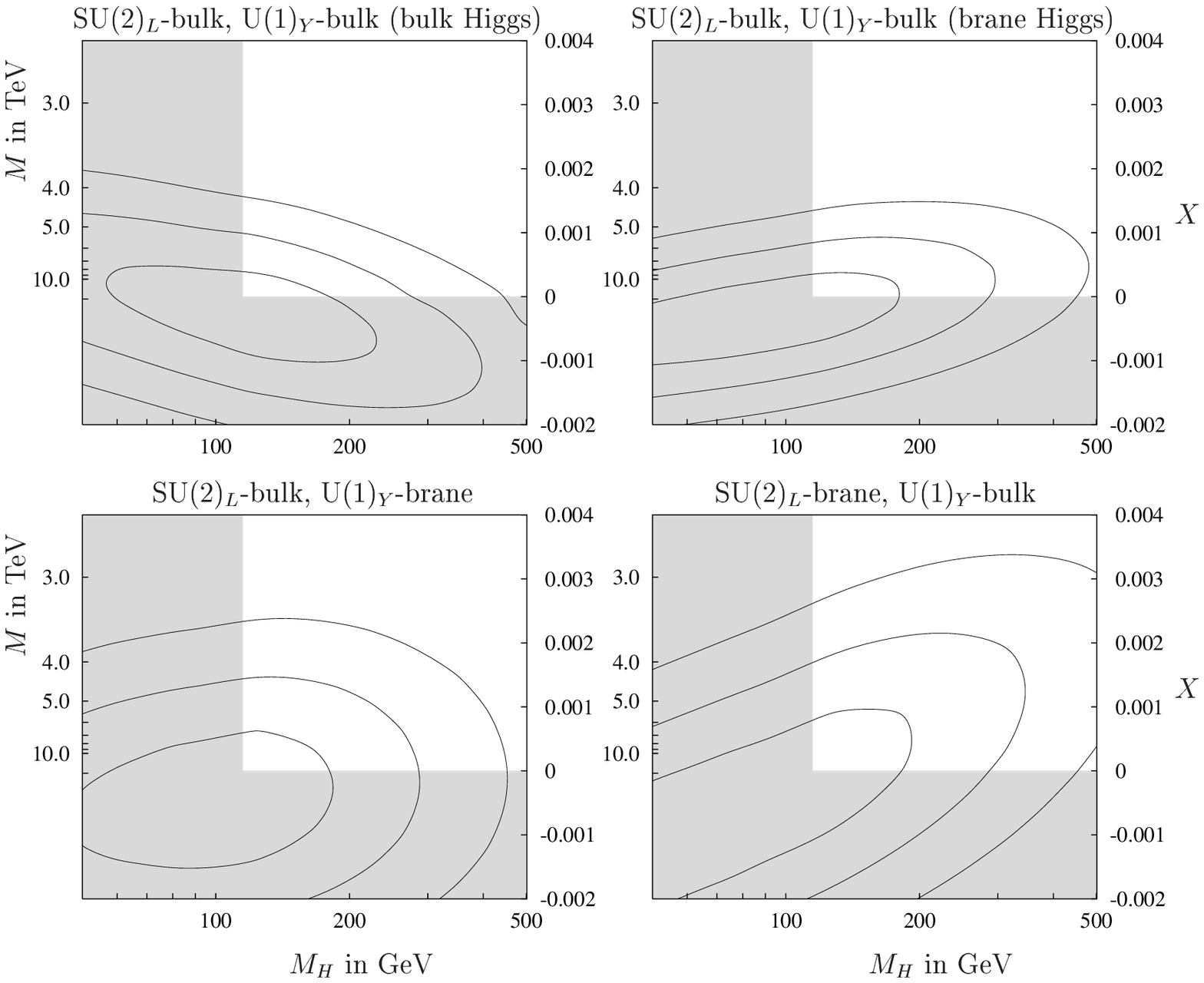}
\mycaption{\label{contoursfulldata} Contours of 
$\Delta \chi^2= 1, \, 4, \, 9$ (eq.   \ref{deltachibound})
derived  from  the  combined  analysis of   LEP2 data  and electroweak
precision measurements.   The  shaded  regions of  the parameter space 
correspond  to $m_H < 114$~GeV  and/or $M^2 < 0$.}
\end{figure}

\subsection{Combined Bounds on the compactification scale $M$}\label{Summary}

The 2$\sigma$ bounds   on $M$ found  from a  one-parameter  fit to the
combined LEP2 data  are listed in Table~\ref{fermiondata}.  The bounds
range  from  3.5~TeV for  the brane-bulk   model  to  5.9~TeV  for the
bulk-bulk  model  with a bulk   Higgs. Furthermore, including also the
electroweak precision measurements, the  bounds range from 4.8~TeV for
the bulk-brane model  to 6.9~TeV for the  bulk-bulk model with a brane
Higgs.   For the bulk-bulk models  our  results agree with the results
of~\cite{CL}.  The best fit values always lie in the unphysical region
$X < 0$.

For muon-pair, tau-pair  and hadron  production (including asymmetries
and heavy-quark data), where ZFITTER  can be used  to calculate the SM
predictions, we  have    also performed a   multi-parameter   fit. The
resulting    $\Delta      \chi^2$   contours      are     shown     in
Fig.~\ref{contoursfulldata}.    The  slight  distortions   from smooth
contours  are due to  a discontinuity in   hadronic cross sections and
asymmetries   found   in   ZFITTER  version   6.36  at    $\sqrt{s}  =
m_{t}$~\cite{ZFDistortions}. The corresponding 2$\sigma$ bounds on $M$
are listed in  Table~\ref{correlationsfulldata} along  with the bounds
from  the  corresponding one-parameter  fit.  The correlations between
$m_H$,  $m_t$,  and  $M$ are similar  to what has  been found from the
precision observables alone in Section~\ref{LEP1}. They  are  weak  in
the bulk-brane model and most sizable in the brane-bulk model.

A comparison of Table~\ref{fermiondata} and \ref{correlationsfulldata}
shows that  in the  one-parameter  fit Bhabha  scattering and $W^+W^-$
production  increase the combined bounds  by about 0.5~TeV.  Thus, the
bounds   from a multi-parameter  fit  to  all  data,  including Bhabha
scattering and  $W^+W^-$ production  can  be estimated to  lie between
4~TeV for the brane-bulk model and 6~TeV for  the bulk-bulk model with
a brane Higgs.

\begin{table}[t] 
\mbox{}
\vspace{-1cm}
\begin{center} 
\begin{tabular}{c|c|c|c|c} 
\hline\hline 
& \pb{2.4}{-0.05}{ \vspace{-0.2cm} brane-bulk }
& \pb{2.4}{-0.05}{ \vspace{-0.2cm} bulk-brane }
& \pb{2.5}{-0.05}{ \vspace{-0.2cm} bulk-bulk \\ (brane Higgs) }
& \pb{2.5}{-0.05}{ \vspace{-0.2cm} bulk-bulk \\ (bulk Higgs) } \\ \hline \hline  
\pb{3.5}{-0.}{one-parameter fit}
& \pb{2}{-0.}{5.0} & \pb{2}{0.}{4.5}
& \pb{2}{-0.}{6.4} & \pb{2}{0.}{5.5} \\ \hline
\pb{3.5}{-0.}{multi-parameter fit}
& \pb{2}{-0.}{3.6} & \pb{2}{0.}{4.3}
& \pb{2}{-0.}{5.4} & \pb{2}{0.}{5.2} \\ \hline \hline
\end{tabular} 
\end{center} 
\mycaption{\label{correlationsfulldata} 2$\sigma$ bounds on $M$ in 
TeV derived from   the  muon-, tau-,  and  hadron  production  at LEP2
combined with electroweak precision measurements.}
\end{table} 

\setcounter{equation}{0}
\section{Sensitivity at a Linear Collider}\label{NLC}

Having  extracted  the bounds on  the  compactification scale $M$ from
available   data, we will now  estimate  the reach  at a future linear
collider such as  TESLA~\cite{TDR}.  For illustration,  we investigate
both the  potential of the GigaZ option  as well as the sensitivity at
high energy and luminosity.

\subsection{GigaZ option}\label{GigaZ}

At  the $Z$ pole,  the  luminosity goal at TESLA   is $\mathcal{L} = 5
\times 10^{33}$~cm$^{-2}$ s$^{-1}$   which is  sufficient  to  produce
$10^9$ $Z$ bosons in only 50-100 days of running~\cite{TDR}. This will
increase the LEP statistics  by more than an  order of  magnitude. The
most  relevant improvement in testing  the compactification scale will
come  from the  precise  measurement of the  left-right (LR) asymmetry
$A_{\mathrm{LR}}$.  Since  photon exchange and  the exchange of higher
KK modes can be  neglected on the  $Z$ peak, the  LR asymmetry at tree
level can be approximated by
\begin{equation}
\label{alr}
A_{\mathrm{LR}}\ =\ \frac{2\, g_{V (0)}\, g_{A (0)}}
                                  {g_{V (0)}^2\: +\: g_{A (0)}^2} \, , 
\end{equation}
where  $g_{V (0)}$  and $g_{A (0)}$   are the vector  and axial vector
couplings   of  the  electron     to  the    $Z$  boson  given      in
(\ref{masseigenstatelagrangian}). This asymmetry  is very sensitive to
shifts of the weak mixing angle  with respect to  the SM value because
of the small ratio  $g_V/g_A=(1- 4 \sin^2 \theta_W)$. Using
(\ref{thetawhatthroughthetaw}) and (\ref{zeromodecouplings}) in  order 
to express (\ref{alr}) in 
terms  of the input parameters, one  obtains the 5DSM corrections
$\Delta_{A_{\mathrm{LR}}} = \Delta_{A_{e}}$ given in~\cite{MPR}.

With the GigaZ option it will be possible to measure $A_{\mathrm{LR}}$
with an absolute  error of  about $10^{-4}$~\cite{TDR}.  However,  the
uncertainties in the  fine structure constant  and  the $Z$  mass will
each induce  an additional error  of about $10^{-4}$  which  has to be
added   in  quadrature.     Coincidence  of the     measured  value of
$A_{\mathrm{LR}}$ with the SM expectation  would then imply the bounds
on $M$ shown in Table~\ref{GIGAZbounds}.   As can be seen, except  for
the bulk-brane  model, the GigaZ  option  should allow  to improve the
existing bounds by at least a factor~2.

With  excellent $b$-tagging, it will  also be possible to considerably
improve the measurement  of the final  state coupling $A_{b}$ and  the
cross-section ratio $R_b$.  The experimental error for the mass of the
$W$ boson can  also be  reduced  significantly  by a threshold   scan.
However, the sensitivity of these observables to $M$ is small and does 
not allow to explore compactification scales beyond the bounds already
known from available data.

\begin{table}[t] 
\mbox{}
\vspace{-1cm}
\begin{center} 
\begin{tabular}{c|c|c|c|c} 
\hline\hline 
& \pb{2.4}{-0.05}{ \vspace{-0.2cm} brane-bulk }
& \pb{2.4}{-0.05}{ \vspace{-0.2cm} bulk-brane }
& \pb{2.5}{-0.05}{ \vspace{-0.2cm} bulk-bulk \\ (brane Higgs) }
& \pb{2.5}{-0.05}{ \vspace{-0.2cm} bulk-bulk \\ (bulk Higgs) } 
\\ \hline \hline  
\pb{2.0}{-0.}{$A_{\mathrm{LR}}$}
& \pb{2}{-0.}{12.4} & \pb{2}{0.}{6.8}
& \pb{2}{-0.}{14.2} & \pb{2}{0.}{12.4} \\ \hline
\end{tabular} 
\end{center} 
\mycaption{\label{GIGAZbounds} 2$\sigma$ bound on the 
compactification scale $M$ in TeV which can be obtained with the GigaZ
option  at  TESLA if the measured value of $A_{\mathrm{LR}}$ coincides 
with the SM expectation.}
\end{table} 

\subsection{Tests at $\sqrt{s} = 800$ GeV}\label{highenergies}

At high energies, the interference  effects from  the  exchange  of SM
and KK modes completely dominate the  mixing effects as illustrated in
Fig.~\ref{sdependence}.  We  consider    the same    processes  as  in
Section~\ref{LEP2}, except for  $W$-pair production which is not  very
sensitive to an extra dimension since the couplings  of the $W$ bosons
to  higher  KK  modes  in  the   $s$-channel are either  forbidden  or
suppressed.     For     Bhabha     scattering,    an  acceptance  cut, 
$|\cos \vartheta| < 0.9$, is included. Furthermore,       we assume an 
integrated luminosity of 1000~fb$^{-1}$.

\begin{figure}[t]
\begin{center}
\includegraphics[width=16cm]{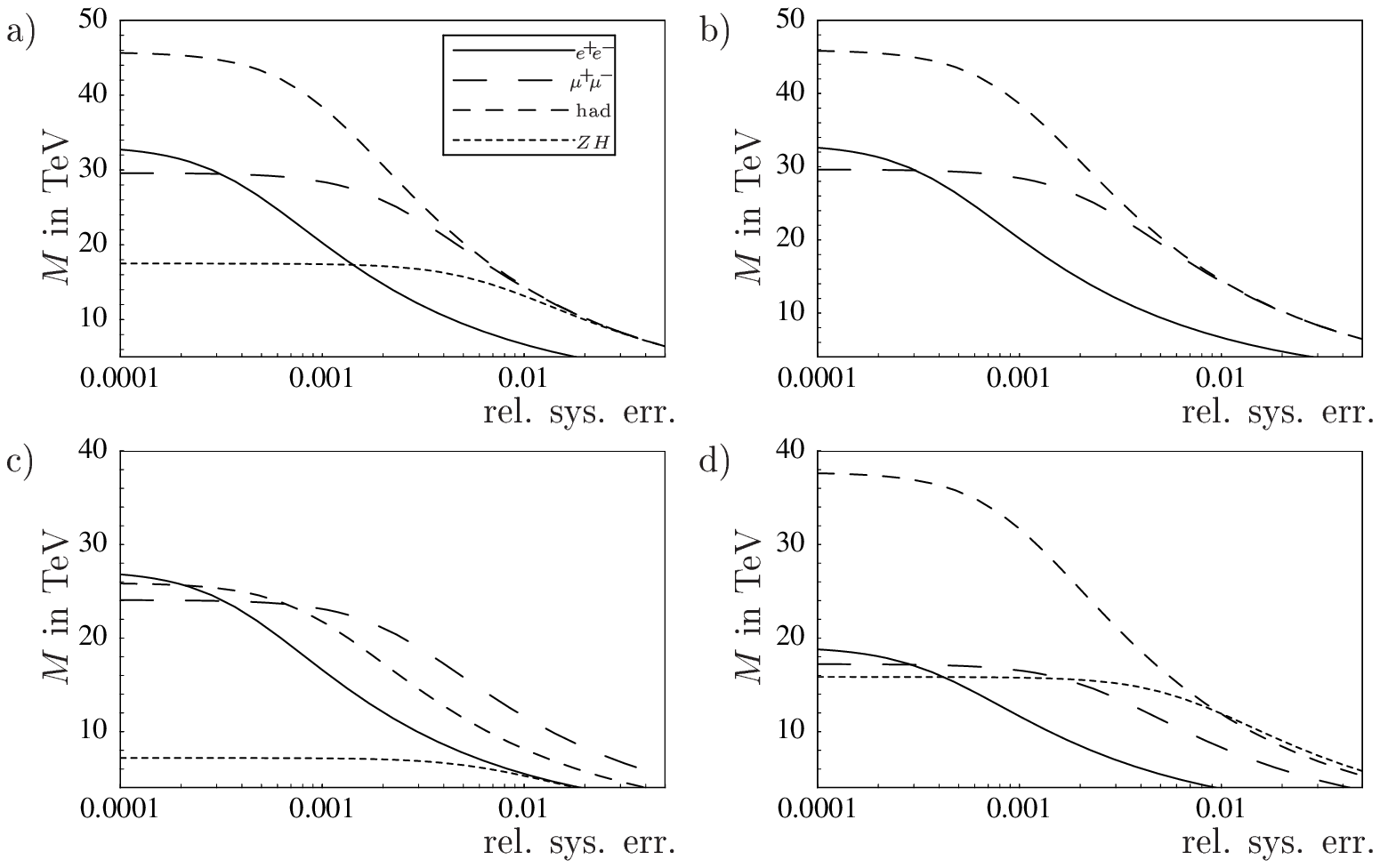}
\end{center}
\mycaption{\label{TESLAvariables} Expected sensitivity to $M$ as   
a function  of    relative systematic  error         at  $\sqrt{s}  =
800$~GeV and   for  an integrated  luminosity  of  1000~fb$^{-1}$: (a)
bulk-bulk model with brane Higgs, (b) bulk-bulk model with bulk Higgs,
(c) brane-bulk model, and (d) bulk-brane model.}
\end{figure}

In  addition,  we also  study  Higgsstrahlung, the  differential cross
section of which is given by
\begin{equation}
\label{diffcrosssectionZH} 
\frac{d \, \sigma (e^+ e^- \to Z \, H)}{d \, \cos \vartheta}\ =\ 
 \frac{s}{512 \pi} \, \lambda^{1/2}(s) \,  
  \lreckig{8 \frac{m_{Z (0)}^2}{s}\: +\: \lambda(s) \lr{1 - \cos^2 \vartheta} } 
\, \lreckig{M_L^2(s) + M_R^2(s)} \, .
\end{equation} 
Here, $\lambda(s) = \! \lr{\!1 - \lr{m_H  + m_{Z (0)}}^2/s} \lr{\!1  - 
\lr{m_H - m_{Z (0)}}^2/s}$   is  the familiar two-particle phase-space 
function, $\vartheta$ is  the  scattering  angle  between the electron
and  the outgoing $Z$, and
\begin{equation}  
M_{\alpha}(s) = \sum_{n=0}^{\infty}  
\lr{ \frac{g^e_{\alpha (n)} \, \, g^{ZH}_{(n)}}{\cos^2 \theta_W}
     \frac{1} {s - m^2_{Z (n)}} } \, . 
\end{equation} 
The couplings $g^e_{\alpha (n)}$ are  defined in (\ref{couplings}) and
$g^{ZH}_{(n)}$ is the effective coupling of the $Z$ modes to the Higgs
boson  given  in Appendix~\ref{higgsstrahlung}.   The integrated cross
section following from (\ref{diffcrosssectionZH}) in  the SM limit can
be found, for example, in~\cite{KKZ}.

The Higgsstrahlung process is  certainly not a primary search  channel
for an  extra dimension because of  the limited experimental accuracy.
For       $\int \! \mathcal{L} \, dt=10^3$~fb$^{-1}$ and at $\sqrt{s}=
800$~GeV, the error on  the  total  cross  section  for    $m_H \simeq
115$~GeV is expected to be~5\%~\cite{TDR}. Nevertheless, this  channel
is interesting for distinguishing between a brane and a bulk Higgs  in
the bulk-bulk model. If the produced Higgs boson is the zero mode of a
bulk field, the   KK     selection    rules  forbid   the     coupling
$H_{(0)}Z_{(0)}Z_{(n)}$  for $n \ge 1$.   Thus, the absence of massive
KK-modes in the s-channel is a clear signal  for a bulk Higgs. In this
case, this  process  is rather SM-like in  contrast  to  a Higgs boson
localized on the brane.  The  phenomenology of 2-Higgs-doublet  models
with a bulk and a brane Higgs has been investigated in~\cite{ABD}.

\begin{figure}[t]
\begin{center}
\includegraphics[width=10cm]{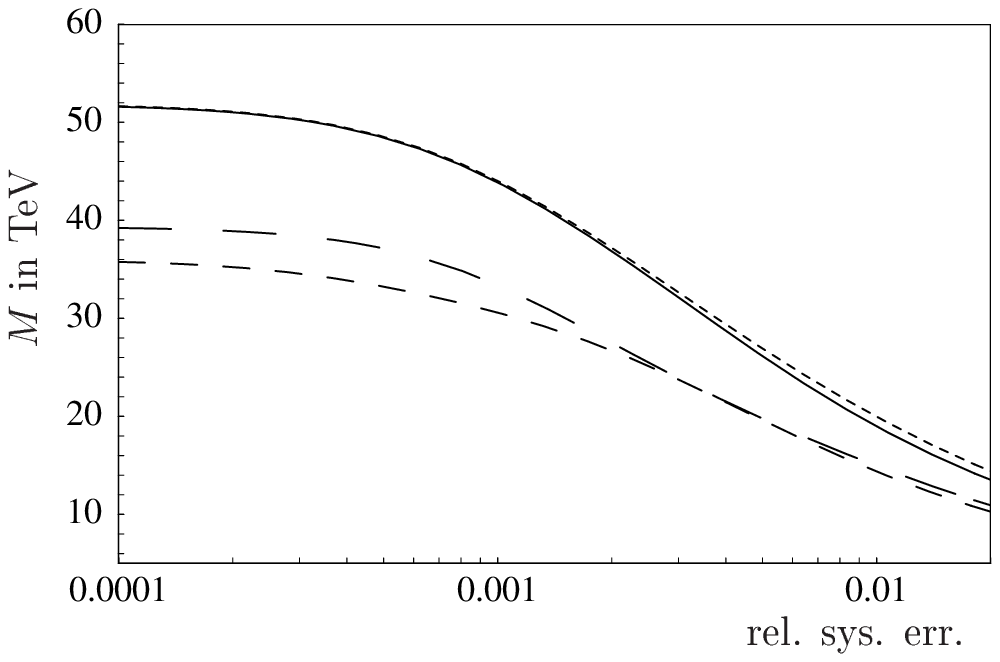}
\end{center}
\mycaption{\label{TESLAcombined} Combined sensitivity to the 
compactification scale $M$ as  a function of relative systematic error
at   $\sqrt{s}  =  800$~GeV  and  for   an   integrated  luminosity of
1000~fb$^{-1}$  in  the models    bulk-bulk with  brane  Higgs  (short
dashed), bulk-bulk with  bulk Higgs (solid), brane-bulk  (dashed), and
bulk-brane (long dashed).}
\end{figure}

In   the following,   the sensitivity  to  the presence   of  an extra
dimension is estimated by requiring
\begin{equation} 
\chi^2 = \sum_{i} \, 
\frac{\lr{ \mathcal{O}_i^{\mr{SM}}\: -\: 
\mathcal{O}_i^{\mr{5DSM}} }^2}{(\Delta \mathcal{O}_i)^2} \, \le \, 4 \, .
\end{equation}
The  statistical errors will be  so  small that the systematic  errors
become decisive~\cite{SR,DB}. Since  the latter are not reliably known
at the present   time,  we  show,  in Fig.~\ref{TESLAvariables},   the
sensitivity  to the  compactification scale $M$  as a  function of the
relative  systematic  error.   At  $\Delta  \mathcal{O}_{\mathrm{sys}}
\stackrel{<}{{}_\sim}  0.001$  the statistical  uncertainty begins  to
dominate and the sensitivity to the compactification scale saturates.

\begin{figure}[t]
\begin{center}
\includegraphics[width=16cm]{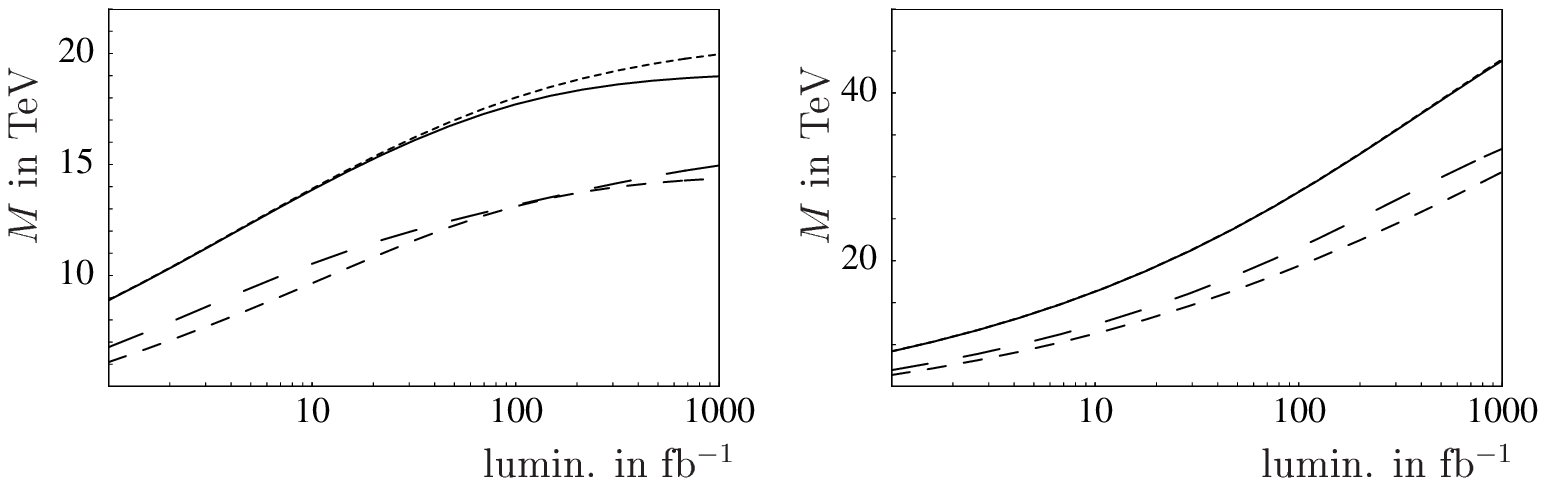}
\end{center}
\mycaption{\label{lumiplot} Search limit as a  
function of the integrated luminosity at $\sqrt{s}~=~800$~GeV
assuming a systematic error of 1.0\%  (left)  and
0.1\% (right). Combination of channels, and models as 
in  Fig.~\ref{TESLAcombined}.}
\end{figure}

The    combined  sensitivity      of  the     search     channels   of
Fig.~\ref{TESLAvariables}          is           presented           in
Fig.~\ref{TESLAcombined}.    For  bulk/brane  (bulk/bulk)  models, the
sensitivity limit increases  from 15~(20)~TeV for systematic errors at
the 1\% level to 35~(50)~TeV for negligible systematic errors.     The
role played by  the statistics is illustrated in  Fig.~\ref{lumiplot},
where the combined sensitivity is plotted  as a function of integrated
luminosity.

In addition to  integrated cross sections, it  is also interesting  to
study the effects of an extra dimension on angular distributions. This
may provide further handles to discriminate between  different models.
For the muon and tau channel, the    angular   distribution   in   the 
bulk-bulk model is almost completely SM-like. However, the  brane-bulk
and   the bulk-brane models lead  to  significant  distortions of  the
angular distribution because of the almost pure U(1)$_Y$ and SU(2)$_L$
nature of the heavy KK modes. In Bhabha  scattering, the  $s$- and the
$t$-channel    are    affected differently    such  that   the angular
distribution is also affected in the bulk-bulk models.

\begin{figure}[t]
\begin{center}
\includegraphics{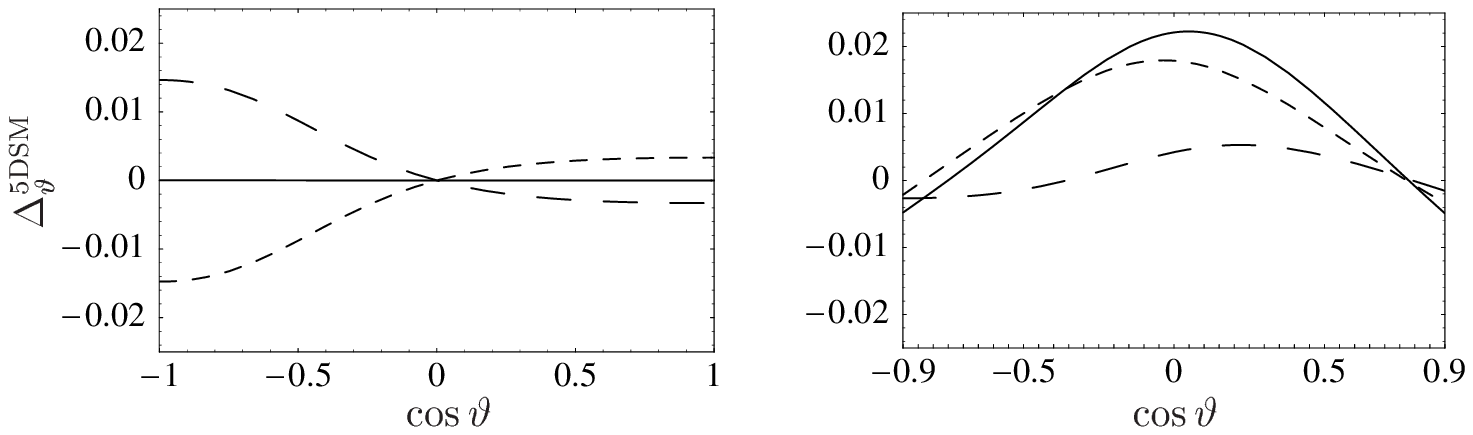}
\end{center}
\mycaption{\label{binnedcrosssectionmuon} Deviation of the
angular distribution $(d \sigma/d \cos \vartheta)/
\sigma_{\mathrm{tot}}$  in muon-pair  production (left) and
Bhabha scattering (right) from  the SM predictions for $M~=~10$~TeV in
the bulk-bulk (solid),   brane-bulk  (dashed),  and  bulk-brane  (long
dashed) models.}
\end{figure}

For illustration,  Fig.~\ref{binnedcrosssectionmuon}  shows the  shift
$\Delta^{\mathrm{5DSM}}_{\mathrm{\vartheta}}$ of $(d  \sigma/d    \cos
\vartheta)/\sigma_{\mathrm{tot}}$ from the SM prediction as defined in
(\ref{generalformofpredictions}). If the angular distributions in  the
muon channel can be measured with a precision  better than 1\% per bin
(using ten  bins), one can probe the compactification scale $M$ beyond
10~TeV   for  the  brane-bulk   and the bulk-brane   model.  In Bhabha
scattering, one can   reach  a similar  scale also  for  the bulk-bulk
models,    while  the  bulk-brane  model is difficult to probe in this 
channel.

\setcounter{equation}{0}
\section{Conclusions}\label{Conclusions}

In this article, we have determined detailed  and robust bounds on the
compactification scale $M$ from  electroweak  precision data and  LEP2
measurements of fermion and W pair production. Our analysis   includes
correlations  of experimental errors in  the data.  Moreover,  besides
one-parameter fits,  we have performed  multi-parameter fits in  order
to   include   correlations   between   the   SM  parameters   and the 
compactification  scale~$M$.     In addition, we  have  estimated  the
sensitivity to  $M$ at  a  future $e^+e^-$ collider such as TESLA. For
this estimate,    we  have  considered  fermion-pair  production   and 
Higgsstrahlung.

If both  the  SU(2)$_L$  and  U(1)$_Y$  fields  are  bulk fields,  the
existing    data imply $M>5.5-6$~TeV     where the range  reflects the
dependence on details   of  the Higgs sector.    If the SU(2)$_L$   or
U(1)$_Y$ fields are confined to the brane where  the fermions live the
bounds are $M > 4$~TeV and  $M > 5$~TeV, respectively. Furthermore, we
have shown that the presence of an extra compact dimension relaxes the
upper bound on the SM Higgs mass from 280~GeV in the  SM (for the data
set  used in   Section~\ref{LEP1})  to 400~GeV    and  330~GeV in  the
brane-bulk and the bulk-bulk model with a brane Higgs, respectively.

At an  $e^+e^-$ linear collider,   the  GigaZ option should  allow  to
increase  the  sensitivity  to $M$  by  a factor~2   in almost  all 5D
models. At $\sqrt{s} =  800$~GeV and for  an integrated luminosity  of
1000~fb$^{-1}$ the discovery  potential will  crucially depend on  the
control of systematic errors.  For a systematic  uncertainty of 1\% in
each search channel, one will be able to reach compactification scales
in the range 15-20~TeV.  For systematic uncertainties smaller than the
statistical uncertainties, the sensitivity limit is estimated to be in
the range $M =$ 35-50~TeV.

Finally,   for   a   sufficiently   low compactification     scale, $M
\stackrel{<}{{}_\sim}   10$~TeV,  Higgsstrahlung        and    angular
distributions of 2-fermion final  states  can be used to  discriminate
between different 5D models. In particular, Higgsstrahlung can be used
to distinguish brane from bulk Higgs bosons.

\subsection*{Acknowledgements}

This work was supported in part by the Bundesministerium f\"ur Bildung
und  Forschung (BMBF, Bonn,   Germany)  under contract 05HT1WWA2,  the
Studienstiftung des  deutschen  Volkes,    and  PPARC   grant   number
PPA/G/O/2000/00461.     We  wish  to  thank   the participants  of the
Heidelberg theory colloquium for  an inspiring discussion and G. Quast
for his help with ZFITTER.

\newpage

\def\theequation{\Alph{section}.\arabic{equation}}
\begin{appendix}

\setcounter{equation}{0}
\section{Kaluza--Klein $W_{(0)}W_{(0)}Z_{(n)}$ and $W_{(0)}W_{(0)}\gamma_{(n)}$ 
couplings}\label{wpair}

\begin{figure}[t]
\begin{center}
\includegraphics{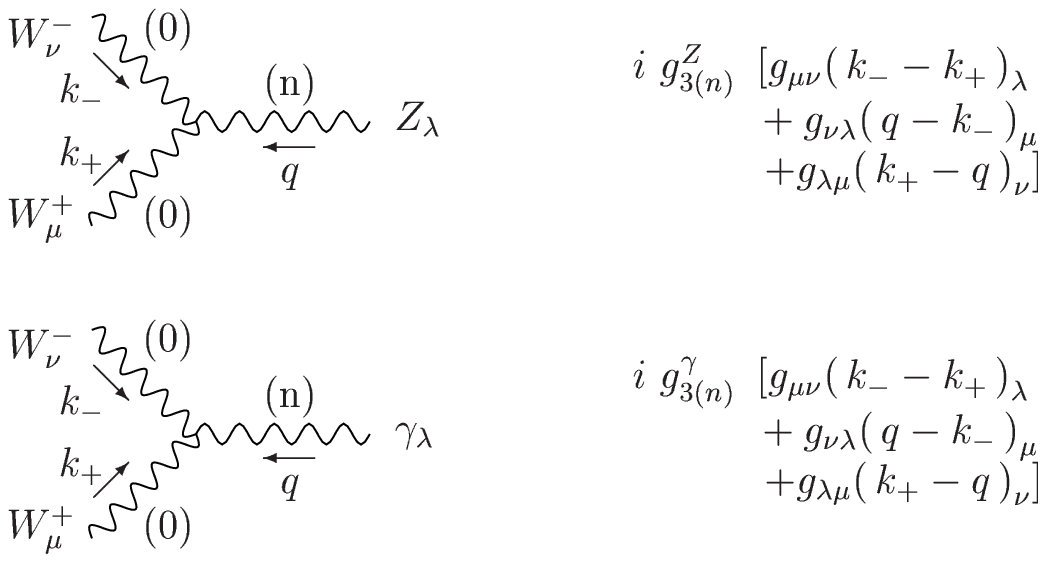}
\end{center}
\mycaption{\label{Feynrules} Triple gauge boson 
couplings.  The numbers in parenthesis denote the KK mode numbers.}
\end{figure}

Here, we present the Feynman rules for the triple gauge boson vertices
shown   in  Fig.~\ref{Feynrules}.   In  the  gauge   basis,  only  the
$W^+_{(0)} W^-_{(0)}  Z_{(0)}$ and $W^+_{(0)}  W^-_{(0)} \gamma_{(0)}$
vertices exist  while the  vertices $W^+_{(0)} W^-_{(0)}  Z_{(n)}$ and
$W^+_{(0)} W^-_{(0)} \gamma_{(n)}$ with $n  \ge 1$ are forbidden by KK
selection   rules~\cite{MPR,DMN}.    However,   in the mass eigenstate
basis,       couplings to   heavy  KK  states   are induced   by   the
diagonalization of the gauge-boson mass  matrix.  Below, we give these
couplings to   first  order in   $X$.   To this  order  the  zero-mode
couplings are unaffected, that is
\begin{equation} 
g^{Z}_{3 (0)} = g \, \cos \theta_{W} \, , \quad \quad 
g^{\gamma}_{3 (0)} = e \, 
\end{equation}
with $g$ from (\ref{gaugecoupl}). For the higher modes ($n \ge 1$), one
gets 
\begin{equation} 
g^{Z}_{3 (n)} = \sqrt{2} \, e \, \frac{\hat{c}_W}{\hat{s}_W} \, 
\lr{\hat{s}_W^2 - \hat{c}_W^2} \, 
                            s^2_{\beta} \, \frac{3}{n^2 \pi^2} \, X \, 
\end{equation}
\begin{equation} 
g^{\gamma}_{3 (n)} = - \sqrt{2} \, e \, \hat{c}^2_W \, s^2_{\beta} \, 
                                           \frac{6}{n^2 \pi^2} \, X \, 
\end{equation}
in the bulk-bulk model,
\begin{equation} 
g^{Z}_{3 (n)} = \sqrt{2} \,  e \, \hat{c}_W \, 
                                           \frac{3}{n^2 \pi^2} \, X \, 
\end{equation}
in the brane-bulk model, and 
\begin{equation} 
g^{Z}_{3 (n)} = - \sqrt{2} \, e \, \frac{\hat{c}^2_W}{\hat{s}_W} \, 
                                       \frac{3}{n^2 \pi^2} \, X \,  
\end{equation}
in the bulk-brane model. In the latter two models, the $\gamma_{(n)}$ 
modes  for $n \ge 1$  are absent.

\setcounter{equation}{0}
\section{Kaluza--Klein $H_{(0)}Z_{(0)}Z_{(n)}$ couplings}\label{higgsstrahlung}

In the bulk-bulk model with a brane Higgs only, the $HZZ$ coupling can
be derived in the gauge basis from
\begin{equation}
\label{ZHHlagrangian}
\begin{split}
\la_{\mr{HZZ}}(x) 
& = \frac{1}{4} \frac{g^2 v}{c_W^2} \, h \, 
    \lr{Z_{(0)}^{\mu} + \sum_{n=1}^{\infty} \sqrt{2} \, Z_{(n)}^{\mu}}^2 \\
& = \frac{g}{2} \, \frac{m_{Z (0)}}{c_W} 
    \left(1 - \frac{\Delta_Z}{2} X \right) \, h \, 
    \lr{Z_{(0)}^{\mu} + \sum_{n=1}^{\infty} \sqrt{2} \, Z_{(n)}^{\mu}}^2 \, ,
\end{split}
\end{equation}
where $h$ denotes  the Higgs field on the  brane and $v$ its VEV.  The
second relation follows from (\ref{deltazdef}). In the bulk-bulk model
with  a bulk Higgs field, the KK selection rules  forbid the couplings
of two zero modes to  higher modes. Moreover,   the  gauge eigenstates
coincide with the mass eigenstates. Thus, for  zero-mode final states,
Higgsstrahlung   is  described    by  the same $H_{(0)}Z_{(0)}Z_{(0)}$
vertex as in the SM. In  the brane-bulk  or  bulk-brane   model,   the
$Z_{(n)}^{\mu}$ tower coincides   with the U(1)$_Y$ or  SU(2)$_L$   KK
modes  for $n \ge 1$, respectively. Because a brane Higgs field breaks
momentum  conservation  in  the extra  dimension,   no selection rules 
exist.

\begin{figure}[t]
\begin{center}
\includegraphics{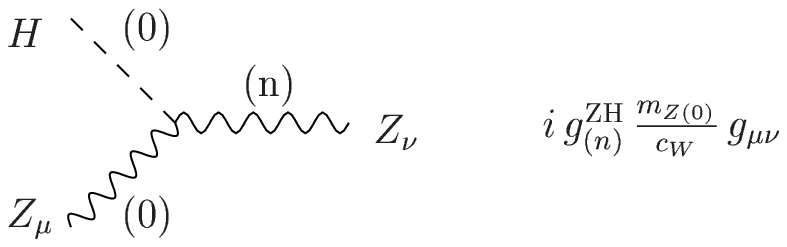}
\end{center}
\mycaption{\label{ZHHvertices} The $HZZ$ vertex. The numbers in parenthesis  
specify the KK~modes.}
\end{figure}

In the mass  eigenstate  basis, the Lagrangian   (\ref{ZHHlagrangian})
leads to the vertex shown in Fig.~\ref{ZHHvertices}. In  summary,  the 
effective couplings are given by
\begin{equation} 
g^{\mr{ZH}}_{(0)} = g
\quad \quad \mr{and} \quad \quad g^{\mr{ZH}}_{(n \ge 1)} = 0
\end{equation}
in the bulk-bulk model with a bulk Higgs,
\begin{equation} 
\begin{split}
g^{\mr{ZH}}_{(0)} \, & = \, g \, 
\left[1 \, - \, \left(2 + \frac{\Delta_Z}{2} \right)\, X \right] \, , \\[2ex]
g^{\mr{ZH}}_{(n \ge 1)} \, & = \, \sqrt{2} \, g \, 
\lreckig{1 \, - \, \lr{1 + \frac{3}{2 \, n^2 \pi^2} + \frac{\Delta_Z}{2}} \, X} \, ,
\end{split}
\end{equation}
in the bulk-bulk model with a brane Higgs,
\begin{equation} 
\begin{split}
\label{ZHcouplingbranebulk}
g^{\mr{ZH}}_{(0)} \, & = \, g \, 
\left[ 1 \, - \, \left( 2 \, \hat{s}_W^2 + \frac{\Delta_Z}{2} \right) \, X \right]\, , \\[2ex]
g^{\mr{ZH}}_{(n \ge 1)} \, & = \, \sqrt{2} \, s_W \,  g \, 
\lreckig{1 \, - \, \lr{\hat{s}_W^2 + \lr{3 \, \hat{s}_W^2 \, - \, 2} 
                       \frac{3}{2 \, n^2 \pi^2} + \frac{\Delta_Z}{2}} \, X} \, ,
\end{split}
\end{equation}
in the brane-bulk model, and 
\begin{equation} 
\begin{split}
g^{\mr{ZH}}_{(0)} \, & = \, g \, 
\left[1 \, - \, \left( 2 \, \hat{c}_W^2 + \frac{\Delta_Z}{2} \right) \, X \right] \, , \\[2ex]
g^{\mr{ZH}}_{(n \ge 1)} \, & = \, \sqrt{2} \, c_W \,  g \, 
\lreckig{1 \, - \, \lr{\hat{c}_W^2 + \lr{3 \, \hat{c}_W^2 \, - \, 2} 
                      \frac{3}{2 \, n^2 \pi^2} + \frac{\Delta_Z}{2}} \, X} \, ,
\end{split}
\end{equation}
in bulk-brane model.    The factor $\sqrt{2}$ in  $g^{\mr{ZH}}_{(n \ge
1)}$ is the usual enhancement of couplings between higher KK modes and
brane fields. The factors $s_W$ and $c_W$ reflect the fact that for $n
\ge 1$ $Z_{(n)}^{\mu}$ is mainly $B^{\mu}_{(n)}$ or $A^{3 \mu}_{(n)}$,
respectively.

\end{appendix}

\end{document}